\documentclass[conference]{IEEEtran}
\usepackage{amsfonts}

\begin{document}

\title{Recursive Secret Sharing for Distributed Storage and Information Hiding}

\author{\IEEEauthorblockN{Abhishek~Parakh}
\IEEEauthorblockA{Computer Science Department\\ Oklahoma State University\\
Stillwater, OK 74078 USA\\
Email: parakh@cs.okstate.edu}
\and
\IEEEauthorblockN{Subhash~Kak}
\IEEEauthorblockA{Computer Science Department\\ Oklahoma State University\\
Stillwater, OK 74078 USA\\
Email: subhashk@cs.okstate.edu}}

\maketitle

\begin{abstract}
This paper presents a recursive computational multi-secret sharing technique that hides $k-2$ secrets of size $b$ each into $n$ shares of a single secret $S$ of size $b$, such that any $k$ of the $n$ shares suffice to recreate the secret $S$ as well as all the hidden secrets. This may act as a steganographic channel to transmit hidden information or used for authentication and verification of shares and the secret itself. Further, such a recursive technique may be used as a computational secret sharing technique that has potential applications in secure and reliable storage of information on the Web, in sensor networks and information dispersal schemes. The presented technique, unlike previous computational techniques, does not require the use of any encryption key or storage of public information.
\end{abstract}

\section{Introduction}
An information theoretically secure $k$-out-of-$n$ secret sharing technique used to share a secret of size $b$ requires a total storage space of size $b\cdot n$. Since, $k-1$ shares do not reveal any information about the secret, such techniques use $k-1$ random elements of size $b$ in order to create the shares. In this paper, we propose that these random elements be replaced with certain hidden information that may serve as a steganographic channel. Note that if a secret sharing scheme uses $k-1$ random elements then $k-1$ is the upper limit on the number of secrets that can be hidden. We hide $k-2$ secrets which is near optimal.

If user $A$ transmits a secret message to user $B$ over a public channel, he may divide the message into several pieces (possibly redundant) and send the pieces on parallel channels, such that an eavesdropper may need to compromise at least $k$ out of $n$ channels to retrieve the message. $B$ upon receiving the pieces may reconstruct the message and authenticate it using the signed hash of the message that $A$ sends to $B$. Transmission of this signature is an additional burden on the network. In the proposed scheme, $A$ may hide the signature within the pieces of the message that is transmitted.

Information dispersal schemes for distributed storage networks primarily use computational secret \cite{ref22,ref9} sharing schemes. In general, in a computational secret sharing scheme an encryption key is used to encrypt the secret that is to be securely stored/transmitted. The encrypted message is then divided into several (possibly redundant) pieces. The key is divided into shares using conventional secret sharing techniques and these shares are stored along with the pieces of the encrypted message, as an overhead \cite{ref7,ref8,ref9}.

In a multiparty scenario, such as in secret sharing, the hidden information may be used as a means of authentication of the share (and of reconstructed secret), thus provide cheating detection. For example, the dealer may hide a ``specially" chosen message in the shares of the secret and distribute the hash of this message to all the players along with the shares. The players may later reconstruct the secret and the hidden message, find the hash of the hidden message and verify it against the hash they have.

The presented scheme may be used as a multi-secret sharing scheme that uses Shamir's secret sharing scheme as its building block and encodes $k-2$ additional secrets within the shares of the message originally intended to be shared. And the scheme may be used as a computational secret sharing scheme, effectively resulting in smaller shares, by dividing a secret into smaller pieces and then simulating a multi-secret sharing scheme. Since the proposed algorithm generates shares on the order of size of secrets encoded, smaller pieces will give rise to smaller shares. Moreover, the proposed scheme does not require any encryption key.

An efficient method for sharing multiple secrets with security based on assumption of hardness of discrete logarithm problem is presented in \cite{ref26}. Whereas \cite{ref23} proposes a scheme based on systematic block codes and \cite{ref24} propose schemes based on Shamir's secret sharing scheme but require a large amount of side information to be stored as public knowldege and further \cite{ref23,ref24,ref30} attempt to maintain ideal security. Other schemes \cite{ref28,ref29} focus at improving efficiency of computations involved in share creation and secret reconstruction rather than space and transmission efficiency.

%For example, in a $k$-out-of-$n$ scheme, computational secret sharing techniques divide the secret of size $b$ into shares of size $\frac{b}{k}$ plus a key of size $m$, each. Hence, the total storage space required is $(\frac{b}{k}+m)\cdot n$ compared to $b\cdot n$ in traditional schemes. This represents a reduction in storage space if $(\frac{b}{k}+m)<b$, which is generally true.

%In order to improve space efficiency, computational secret sharing techniques have been developed \cite{ref2,ref3,ref4, ref10} in which a symmetric key is used to encrypt the original secret and the key is split into shares using conventional methods of secret sharing. The encrypted secret is divided into pieces to which redundancy is added by the use of block error correction techniques \cite{ref7,ref8,ref9}. This leads to a $n$-fold increase in key size, pieces of which have to be stored with every share of the encrypted secret, hence becoming a overhead. Moreover, this reduction in storage is achieved by relaxing the security requirements, since the computational security is weaker than information theoretic security \cite{ref9}.

%Consequently, we ask whether it is possible to improve space efficiency of secret sharing techniques while maintaining information theoretic security. This question is answered, herein, in the affirmative by proposing a secret sharing scheme that encodes $k-1$ secrets in $n$ shares compared to conventional methods of encoding 1 secret in $n$ shares, thus increasing space efficiency and maintaining information theoretic security.

In an earlier paper \cite{ref5,ref6}, a 2-out-of-2 ($k=2$ and $n=2$) recursive scheme for secret sharing was proposed. In this method, if $k$ secrets are chosen such that they double in size, then all of the smaller secrets can be recursively stored in the shares of larger secrets, so that two shares of size $2^m$ can encode $2^{m+1}-1$ bits of information. For example, if we are to share 3 secrets $s_1=1$, $s_2=01$, and $s_3=1011$, then the two shares for $s_1$ would be $D_{{s_1}1}=0$ and $D_{{s_1}2}=1$; where exclusive-OR operation is used for secret reconstruction. The shares of $s_1$ can be used to create two shares of $s_2$ as follows: $D_{{s_2}1}=D_{{s_1}1}0=00$ and $D_{{s_2}2}=0D_{{s_1}2}=01$. Here $D_{{s_1}1}0$ denotes concatenation of share 1 of secret $s_1$ with 0; and $0D_{{s_1}2}$ denotes concatenation of 0 with share 2 of secret $s_1$, and so on. Similarly, we can recursively use the shares of $s_2$ to create the shares of $s_3$: $D_{{s_3}1}=D_{{s_2}1}10=0010$ and $D_{{s_3}2}=10D_{{s_2}2}=1001$. As a result, the final two shares for all the three secrets are 0010 and 1001. Consequently, using 8 bits of shares we have encoded 7 bits of secrets. This is in comparison with conventional methods that would require a total 14 bits of shares.

The above efficiency increase is obtained as a tradeoff against security of the scheme. A non-recursive scheme would require 7 bits for each share but the recursive scheme requires 4 bits per share, and a player only needs to determine 4 bits to break the scheme. However, in practice secrets are thousands of bits long. For example, a secret of 1048 bits length would be encoded in approximately 1024 bits per share, and would still require $2^{1024}$ combinations to break. This may be sufficient for many cases.

\section{Space efficient secret sharing}

We propose a method to hide $k-2$ secrets of size $b$, within the shares of a secret $S$ of size $b$, using a $(k,n)$ modified Shamir's secret sharing scheme. The secret is divided into $n$ shares using modified Shamir's secret sharing scheme such that any $k$ of them can be brought together for reconstruction.

\noindent\textit{\textbf{Algorithm 1 (Modified Shamir's secret sharing scheme)}}
\begin{enumerate}
  \item Choose a prime $p$, $p>max(S,n)$, where $S$ is the secret.
  \item Choose $k-1$ random numbers $y_1$, $y_2$, ..., $y_{k-1}$, uniformly and independently, from the field $\mathbb{Z}_p$.
  \item Map these random numbers $y_i$s as $y$ coordinates of points: $(i,y_i)$, for all $1\leq y_i\leq (k-1)$.
  \item Map the secret $S$ as point $(0,S)$.
  \item Using $k$ points $(i,y_i)$, for all $1\leq y_i\leq (k-1)$ and $(0,S)$ interpolate a polynomial $p(x)$ of degree $k-1$ modulo prime $p$.
  \item Sample $p(x)$ at $n$ points $D_i=p(i)$, $k\leq i\leq k+n-1$ such that the shares are given by $(i,D_i)$.
\end{enumerate}

The reconstruction procedure for the secret follows the conventional method \cite{ref1}.

Now consider $k-2$ secrets $s_1s_2...s_{k-2}$, $s_i\in\mathbb{Z}_p$ for all $1\leq i\leq (k-2)$ or pieces of a larger message. Therefore our task is to recursively hide $s_i$'s within the shares of secret $S$. Further, we use the notation $y_{lm}$ to denote the $y$-coordinates of points. Here the first subscript $l$ is the index of the step in the recursive process and subscript $m$ is index of share to which that $y$-coordinate belongs to. For example, the $y$-coordinate of share 3 in the $5^{th}$ recursion is written $y_{53}$.

The proposed algorithm works as follows - randomly and uniformly choose a number $y_{11}$ and map it as point $(1,y_{11})$. Using $(0,s_1)$ and $(1,y_{11})$ interpolate $1^{st}$ degree polynomial $p_1(x)$. Sample $p_1(x)$ at two points $y_{21}=p_1(x=2)$ and $y_{22}=p_1(x=3)$. Now map the sampled points as $(1,y_{21})$ and $(2,y_{22})$. Using the next piece as point $(0,s_2)$ and the newly generated points $(1,y_{21})$ and $(2,y_{22})$ interpolate $2^{nd}$ degree polynomial $p_2(x)$. Evaluate $p_2(x)$ at 3 points $y_{31}=p_2(x=3)$, $y_{32}=p_2(x=4)$, and $y_{33}=p_2(x=5)$. We then use these 3 points as y-coordinates for $x$=1, 2, 3 and along with the third piece of the message as point $(0,s_3)$ interpolate $3^{rd}$ degree polynomial $p_3(x)$. We continue this process until we have used all the pieces and reached $(0,s_{k-2})$ and interpolated $(k-2)^{th}$ degree polynomial $p_{k-2}(x)$. We then sample $p_{k-2}(x)$ at $k-1$ points $y_{(k-1)1}=p_{k-2}(k-1)$, $y_{(k-1)2}=p_{k-2}(k)$, $y_{(k-1)3}=p_{k-2}(k+1)$, ..., $y_{(k-1)(k-1)}=p_{k-2}(2k-3)$.

Mapping these $k-1$ samples as points $(1,y_{(k-1)1})$, $(2,y_{(k-1)2})$, ..., $(k-1,y_{(k-1)(k-1)})$ along with $(0,S)$ construct a $(k-1)^{th}$ degree polynomial $p_{k-1}(x)$. We can now sample $p_{k-1}(x)$ at $n$ points such that any $k$ points would reconstruct the secret and the hidden information.

The process of share creation and information hiding is formally described in Algorithm 2.

\noindent\textit{\textbf{Algorithm 2 - Dealing Phase}}

\begin{enumerate}
  \item Consider $k-2$ secrets $s_i\in\mathbb{Z}_p$, $1\leq i\leq (k-2)$.
  \item Choose prime $p=max(s_i,S)$, for all $1\leq i\leq k-2$.
  \item Randomly and uniformly choose a number $y_{11}\in\mathbb{Z}_p$ and map it as point $(1,y_{11})$.
  \item Do for $1\leq i\leq (k-2)$
  \begin{enumerate}
    \item Interpolate points $(0,s_i)$ and $(j,y_{ij})$, for all $1\leq j\leq i$ to generate a $i^{th}$ degree polynomial $p_i(x)$.
    \item Sample the polynomial $p_i(x)$ at $i+1$ points: $y_{(i+1)j}=p_i(j+i)$, for all $1\leq j\leq (i+1)$.
    \item Map the $i+1$ points as: $(j,y_{(i+1)j})$, for all $1\leq j\leq (i+1)$.
  \end{enumerate}
  \item Interpolate points $(0,S)$ and $(j,y_{(k-1)j})$, for all $1\leq j\leq (k-1)$ to generate $(k-1)^{th}$ degree polynomial $p_{k-1}(x)$.
  \item Sample $p_{k-1}(x)$ at $n$ points to generate $n$ shares: $(i,p_{k-1}(i))$, for all $k\leq i\leq k+n-1$.
\end{enumerate}

\noindent\textit{\textbf{Algorithm 2 - Reconstruction Phase}}

\begin{enumerate}
  \item Interpolate any $k$ shares to generate $(k-1)^{th}$ degree polynomial $p_{k-1}(x)=S+a_1x+a_2x^2+...+a_{k-1}x^{k-1}$.
  \item Evaluate $S=p_{k-1}(0)$.
  \item Do for $i=k-2$ down to 1
  \begin{enumerate}
      \item Map the coefficients of polynomial $p_{i}(x)$ as points: $(j,a_j)$, for all $(i+1)\leq j\leq 2(i+1)$.
      \item Interpolate $(j,a_j)$, for all $(i+1)\leq j\leq 2(i+1)$, to generate polynomial $p_i(x)$ of degree $i$.
      \item Evaluate $s_i=p_i(0)$.
  \end{enumerate}
\end{enumerate}

\textit{Security of the proposed method}: Algorithm 2 works by repetitive application of Algorithm 1. The first iteration of the algorithm is a direct application of $(2,2)$ Shamir's secret sharing scheme. It uses a polynomial of degree 1 and generates two shares for the first secret $s_1$ of the message. These two shares may be viewed as random numbers, such that given any number $r\in\mathbb{Z}_p$, $Pr(r=y_{21})=Pr(r=y_{22})=\frac{1}{p}$. They are then used to create a quadratic equation along with the second secret $s_2$ mapped at $x$=0 (the free term of the equation). This quadratic equation is then sampled at 3 points to generate 3 shares of $s_2$. These three shares are then used as random points to generate a $4^{th}$ degree equation and encode $s_3$ and so on, until we have encoded all the $k-2$ pieces and generated $k-1$ shares. These $k-1$ shares are then used as points along with secret $S$ at $x$=0 to generate a polynomial of degree $k-1$, which can then be sampled at $n$ points to create the final shares. These final shares have the shares of the smaller pieces hidden within them. The security of the protocol is predicated upon the random and uniform choice of the first coefficient $y_{11}$.

\textbf{\textit{Example.}} Suppose we want to hide 3 secrets $s_1=46$, $s_2=69$, and $s_3=72$ within the shares of a secret $S$=65. Let $k=5$ and $n=7$, i.e. we are to create 7 pieces such that 5 of them must come together to recreate the secret and the hidden message. We execute the algorithm as follows,
\begin{enumerate}
  \item Choose a prime $p=131$.
  \item Randomly and uniformly choose a number $y_{11}\in Z_{131}$, say $y_{11}=102$. Map it as point $(1,102)$.
  \item Interpolate $(0,s_1)=(0,46)$ and $(1,102)$ to generate $p_1(x)=56x+46$.
  \item Sample $p_1(x)$ at two points $x=2,3$: $y_{21}=p_1(2)=27$ and $y_{22}=p_1(3)=83$.
  \item Map these new points as $(1,y_{21})=(1,27)$ and $(2,y_{22})=(2,83)$.
  \item Interpolate $(0,s_2)=(0,69)$, $(1,27)$ and $(2,83)$ to generate $p_2(x)=49x^2+40x+69$.
  \item Sample $p_2(x)$ at three points $x=3,4,5$: $y_{31}=p_2(3)=106$, $y_{32}=p_2(4)=96$ and $y_{33}=p_2(5)=53$.
  \item Map the new points as: $(1,y_{31})=(1,106)$, $(2,y_{32})=(2,96)$ and $(3,y_{33})=(3,53)$.
  \item Interpolate $(0,s_3)=(0,72)$, $(1,106)$, $(2,96)$ and $(3,53)$ to generate $p_3(x)=111x^3+38x^2+16x+72$.
  \item Sample $p_3(x)$ at 4 points $x=4,5,6,7$: $y_{41}=p_3(4)=119$, $y_{42}=p_3(5)=43$, $y_{43}=p_3(6)=98$ and $y_{44}=p_3(7)=33$.
  \item Map the new points as $(1,y_{41})=(1,119)$, $(2,y_{42})=(2,43)$, $(3,y_{43})=(3,98)$ and $(4,y_{44})=(4,33)$.
  \item Interpolate $(0,S)=(0,65)$, $(1,119)$, $(2,43)$, $(3,98)$ and $(4,33)$ to generate $p_4(x)=66x^4+106x^3+72x^2+72x+65$.
  \item Sample $p_4(x)$ at 7 points $x=5,6,7,8,9,10,11$ to create 7 shares: $(5,p_4(5))=(5,2)$; $(6,p_4(6))=(6,40)$; $(7,p_4(7))=(7,63)$; $(8,p_4(8))=(8,130)$; $(9,p_4(9))=(9,50)$; $(10,p_4(10))=(10,37)$ and $(11,p_4(11))=(11,55)$.
\end{enumerate}

Any five out of the seven shares can be interpolated to regenerate the polynomial $p_4(x)$. This polynomial can then be sampled to obtain $S=p_4(0)$. The $y$-coordinates of the samples of $p_4(x)$ at points $x=1,2,3,4$ can the be mapped as points at $x=4,5,6,7$ and then interpolated to reconstruct $p_3(x)$, which can be sampled at $x=0$ to obtain $s_3=p_3(0)$. Polynomial $p_3(x)$ can be sampled at $x=1,2,3$ to obtain $y$-coordinates and map them at $x=3,4,5$. Interpolating these new points we obtain $p_2(x)$ and so on. The pieces of the hidden secrets are retrieved in the reverse order.

\section{Conclusions}
We have proposed a recursive techniques to hide additional information within the shares of Shamir's secret sharing schemes. This hidden information may be used for validation of shares at the time of secret reconstruction. Further it may be looked upon as a way to share large secrets by dividing the secret in smaller pieces and recursively hiding them in the shares.

Such a scheme is useful for secure transmission of information over parallel channels. Suppose the transmitter and receiver share secret identifications.The transmitter can then divide the identification into pieces and recursively encode it into the shares of the message to be sent over parallel lines. Transmission of shares over parallel channels provided implicit security and reliability. Further, the scheme may be used for information dispersal in storage networks.

\textit{\textbf{Future Work:}} It includes implementing a distributed data storage scheme on the Web where different servers store data by creating shares of the data using the proposed scheme. This implicitly prevents any one (compromised) server from having access all the user data \cite{ref31}. Such an idea may be useful in cloud computing, Chord protocol and FreeNets. Issues regarding addressing of data shares on the network need to be investigated.

\section{Acknowledgment}
This research has been partly funded by the Center for Telecommunication and Network Security (CTANS), Oklahoma State University, Stillwater.

\bibliographystyle{IEEEtran}
\bibliography{IEEEabrv,Sh_Recursive}

\begin{thebibliography}{10}
\providecommand{\url}[1]{#1}
\csname url@samestyle\endcsname
\providecommand{\newblock}{\relax}
\providecommand{\bibinfo}[2]{#2}
\providecommand{\BIBentrySTDinterwordspacing}{\spaceskip=0pt\relax}
\providecommand{\BIBentryALTinterwordstretchfactor}{4}
\providecommand{\BIBentryALTinterwordspacing}{\spaceskip=\fontdimen2\font plus
\BIBentryALTinterwordstretchfactor\fontdimen3\font minus
  \fontdimen4\font\relax}
\providecommand{\BIBforeignlanguage}[2]{{%
\expandafter\ifx\csname l@#1\endcsname\relax
\typeout{** WARNING: IEEEtran.bst: No hyphenation pattern has been}%
\typeout{** loaded for the language `#1'. Using the pattern for}%
\typeout{** the default language instead.}%
\else
\language=\csname l@#1\endcsname
\fi
#2}}
\providecommand{\BIBdecl}{\relax}
\BIBdecl

\bibitem{ref22}
P.~Rogaway and M.~Bellare, ``Robust computational secret sharing and a unified
  account of classical secret-sharing goals,'' in \emph{CCS '07: Proceedings of
  the 14th ACM Conference on Computer and Communications Security}.\hskip 1em
  plus 0.5em minus 0.4em\relax New York, NY, USA: ACM, 2007, pp. 172--184.

\bibitem{ref9}
H.~Krawczyk, ``Secret sharing made short,'' \emph{Proceedings of the 13th
  Annual International Cryptology Conference on Advances in Cryptology}, pp.
  136--146, 1994.

\bibitem{ref7}
M.~O. Rabin, ``Efficient dispersal of information for security, load balancing
  and fault tolerance,'' \emph{Journal of the ACM}, vol.~36, no.~2, pp.
  335--348, 1989.

\bibitem{ref8}
J.~Garay, R.~Gennaro, C.~Jutla, and T.~Rabin, ``Secure distributed storage and
  retrieval,'' \emph{Theoretical Computer Science}, pp. 275--289, 1997.

\bibitem{ref26}
L.~Harn, ``Efficient sharing (broadcasting) of multiple secrets,'' \emph{IEE
  Proceedings - Computers and Digital Techniques}, vol. 142, no.~3, pp.
  237--240, May 1995.

\bibitem{ref23}
H.-Y. Chien, J.-K. Jan, and Y.-M. Tseng, ``A practical (t,n) multi-secret
  sharing scheme,'' \emph{IEICE transactions on fundamentals of electronics,
  communications and computer sciences}, vol.~83, no.~12, pp. 2762--2765, 2000.

\bibitem{ref24}
L.-J. Pang and Y.-M. Wang, ``A new (t, n) multi-secret sharing scheme based on
  shamir's secret sharing,'' \emph{Applied Mathematics and Computation}, vol.
  167, no.~2, pp. 840 -- 848, 2005.

\bibitem{ref30}
C.-W. Chan and C.-C. Chang, ``A scheme for threshold multi-secret sharing,''
  \emph{Applied Mathematics and Computation}, vol. 166, no.~1, pp. 1 -- 14,
  2005.

\bibitem{ref28}
M.~Liu, L.~Xiao, and Z.~Zhang, ``Linear multi-secret sharing schemes based on
  multi-party computation,'' \emph{Finite Fields and Their Applications},
  vol.~12, no.~4, pp. 704 -- 713, 2006.

\bibitem{ref29}
M.~H. Dehkordi and S.~Mashhadi, ``New efficient and practical verifiable
  multi-secret sharing schemes,'' \emph{Information Sciences}, vol. 178, no.~9,
  pp. 2262 -- 2274, 2008.

\bibitem{ref5}
M.~Gnanaguruparan and S.~Kak, ``Recursive hiding of secrets in visual
  cryptography,'' \emph{Cryptologia}, vol.~26, pp. 68--76, 2002.

\bibitem{ref6}
A.~Parakh and S.~Kak, ``A recursive threshold visual cryptography scheme,''
  \emph{Cryptology ePrint Archive, Report 535}, 2008.

\bibitem{ref1}
A.~Shamir, ``How to share a secret,'' \emph{Communications of ACM}, vol.~22,
  no.~11, pp. 612--613, 1979.

\bibitem{ref31}
A.~Parakh and S.~Kak, ``Online data storage using implicit security,''
  \emph{Information Sciences}, vol. 179, no.~19, pp. 3323 -- 3331, 2009.

\end{thebibliography}

\end{document}